# Réalisation, Caractérisation et Modélisation d'une Bobine Supraconductrice à Haute Température Critique de type Double Racetrack

Lauro FERREIRA[1,2], Yasmine BAAZIZI[1,2], Simon MEUNIER[1,2], Tanguy PHULPIN[1,2], Tian-Yong GONG[1,2], Loïc QUÉVAL[1,2]

[1] Université Paris-Saclay, CentraleSupélec, CNRS, GeePs, 91192, Gif-sur-Yvette, France.
[2] Sorbonne Université, CNRS, GeePs, 75252, Paris, France.

**RESUME** - Cet article synthétise nos travaux qui portent sur la réalisation, la caractérisation et la modélisation d'une bobine supraconductrice à haute température critique de type double racetrack (DRC). À l'aide d'un système de bobinage développé en interne, une DRC a été bobinée avec du ruban supraconducteur ReBCO. La bobine a d'abord été caractérisée en courant continu, pour obtenir sa caractéristique *IV* et son courant critique. Puis elle a été caractérisée en courant alternatif (60 Hz et 120 Hz), pour obtenir ses pertes en courant alternatif (pertes AC). Ces mesures ont été comparé avec un modèle éléments finis 3D, et montrent un accord raisonnable compte tenu des hypothèses de modélisation. La bobine présente de bonnes performances, n'a subi aucune dégradation lors du bobinage et pourra être utilisée dans le cadre d'études ultérieures. Ceci valide notre procédé de réalisation et notre approche de caractérisation de bobines supraconductrices.

*Mots-clés— Supraconducteur à haute température critique, Bobine Double Racetrack, ReBCO, Caractérisation IV, Pertes en courant alternatif, modèle COMSOL, température cryogénique.*

## 1. Introduction

Les bobines supraconductrices à haute température critique (HTS) sont prometteuses pour des applications telles que les systèmes de stockage d'énergie magnétique (SMES) [1], des machines électriques [2] ou des transformateurs de puissance [3].

Le supraconducteur étant disponible sous forme de rubans (coated conductor), les bobines présentent généralement des géométries spécifiques telles que le single pancake (Fig. 1a) [4], le double pancake (Fig. 1b) [5], le single racetrack (Fig. 1c) [6] et le double racetrack (Fig. 1d) [7].

Parmi ces topologies, les bobines double pancake/racetrack présentent deux avantages. D'abord, elles permettent d'utiliser des longueurs de conducteur plus importantes sans joint résistif. Ensuite, les connecteurs sont situés à l'extérieur, ce qui simplifie la mise en série ou en parallèle des bobines.

Dans cet article, nous présentons nos avancées concernant la réalisation (section 2), la caractérisation (section 3) et la modélisation (section 4) d'une bobine double racetrack (DRC) bobinée avec du ruban supraconducteur 2G ReBCO.

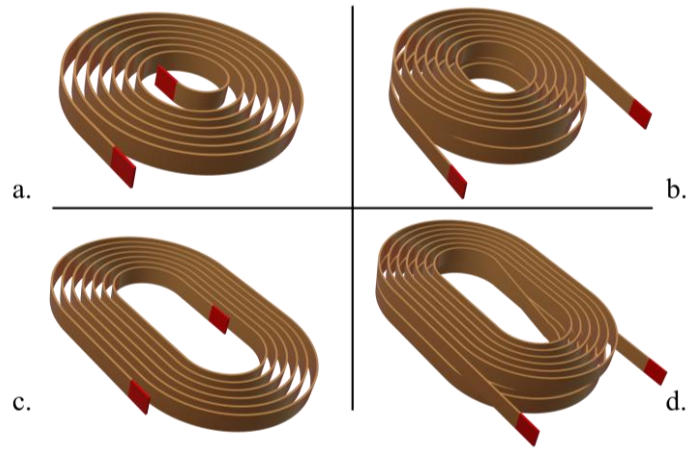

Fig. 1. Types de géométries de bobines. (a) single pancake, (b) double pancake, (c) single racetrack, (d) double racetrack.

## 2. Realisation

Une machine à bobiner a été spécialement mise au point pour réaliser des bobines supraconductrices de différentes tailles et géométries, notamment des bobines single/double pancake/racetrack (Fig. 2).

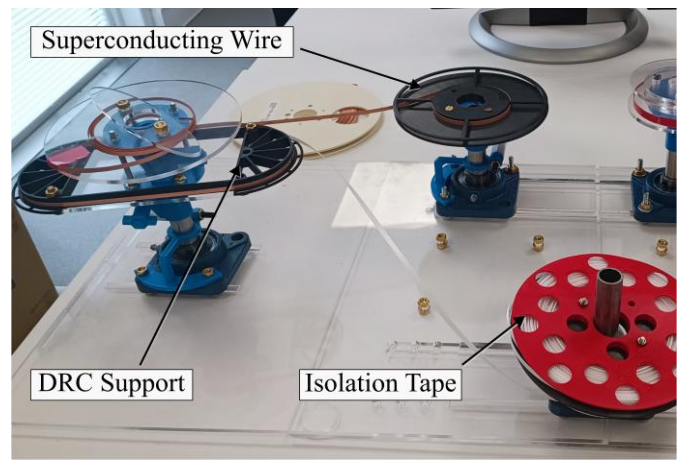

Fig. 2. Machine à bobiner.

Ce système permet un contrôle manuel précis du processus de bobinage, minimisant les efforts mécaniques appliqués au ruban supraconducteur pendant le processus.

A l'aide de cette machine à bobiner, nous avons notamment fabriqué une DRC de 20 spires isolées (10 par couche) avec 12 mètres de ruban HTS ReBCO (SuNAM GdBCO SCN04150) (Fig. 3). Ses caractéristiques sont résumées dans le Tableau I.

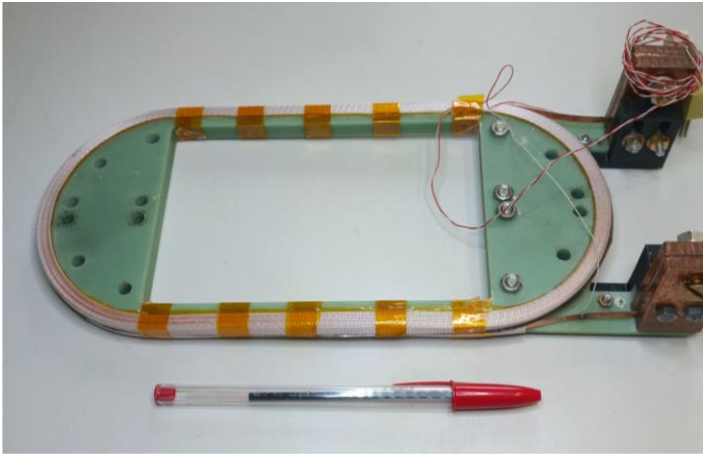

Fig. 3. Bobine double racetrack.

Tableau 1. Caractéristiques de la bobine DRC.

| Paramètres | Spécification |
|---|---|
| Ruban HTS | |
| Matériau | GdBCO (SuNAM SCN04150) |
| Largeur | 4.1 mm |
| Épaisseur du ruban + isolation | 0.14 mm + 1 mm |
| Courant critique datasheet | > 150 A @ 77 K S.F. |
| Courant critique mesuré | 228 A @ 77 K S.F. |
| Indice de loi de puissance mesuré | 44 |
| Support | |
| Matériau | G-10 |
| Dimensions (L × l × h) | 25.5 cm × 12 cm × 1 cm |
| Rayon interne de la section circulaire | 4.5 cm |
| Longueur de la section droite | 14.5 cm |
| DRC | |
| Longueur du ruban HTS | 12 m |
| Nombre de spires | 10 × 2 |
| Matériau des connecteurs | Cuivre |
| Matériau d'isolation | Polypropylène (PP) |
| Courant critique mesuré | 166 A @ 77 K S.F. |
| Indice de loi de puissance mesuré | 46 |

## 3. CARACTERISATION

### 3.1. Caractéristique IV

Les mesures en courant continu (DC) visent à obtenir les caractéristiques *IV*. Ceci est fait à l'aide de la plateforme fort courant 'Gargantua' du laboratoire GeePs. Celle-ci peut fournir jusqu'à 4800 A sous 10 V DC [8]. Le refroidissement est fait en plongeant l'échantillon dans un bain d'azote liquide à 77 K. Pour obtenir la caractéristique *IV*, des échelons de courant sont appliqués tandis que la tension est mesurée par un nanovoltmètre.

Pour référence, nous avons mesuré la caractéristique *IV* d'un court échantillon du ruban utilisé pour le bobinage de la DRC. Les prises de potentiel ont été espacées de 10 cm et pour estimer le courant critique nous avons utilisé le critère de 1 µV/cm. Nous avons mesuré un courant critique de 228 A et un indice de loi de puissance de 44.

Pour la DRC, des prises de potentiel ont été soudées sur le ruban HTS à 1 cm des connecteurs (Fig. 3). Bien que nous reconnaissions la limite d'une telle approche [9], [10], nous avons également utilisé le critère de 1 µV/cm pour estimer le courant critique de la bobine. Nous avons mesuré un courant critique de 166 A et un indice de loi de puissance de 46.

La figure 4 montre la différence entre les caractéristiques *IV* du ruban et de la DRC. La DRC présente une diminution d'environ 30% de son courant critique par rapport au ruban.

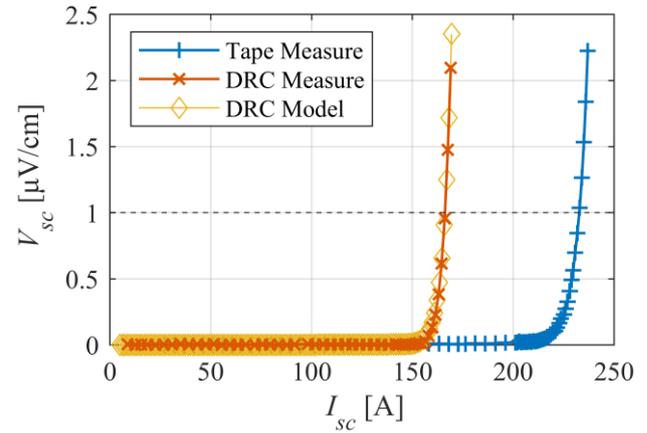

Fig. 4. Caractéristiques *IV* mesurées d'un échantillon court de ruban HTS et de la DRC.

### 3.2. Pertes AC

Les mesures en courant alternatif (AC) visent à obtenir les pertes AC, en utilisant la méthode électrique [11]. Ceci est fait à l'aide de la plateforme 'PAChinko' du laboratoire GeePs (Fig. 5). Celle-ci comprend un générateur de fonction, un amplificateur de puissance, un transformateur abaisseur et une détection synchrone.

Les mesures ont été réalisées à 60 et 120 Hz afin d'éviter les perturbations du réseau à 50 Hz. Sur la figure 6, on trace les pertes AC, notées P, en fonction de l'amplitude du courant alternatif i_peak. En échelle log-log, les pertes sont linéaires par rapport au courant et augmente avec la fréquence, en accord avec la littérature [12].

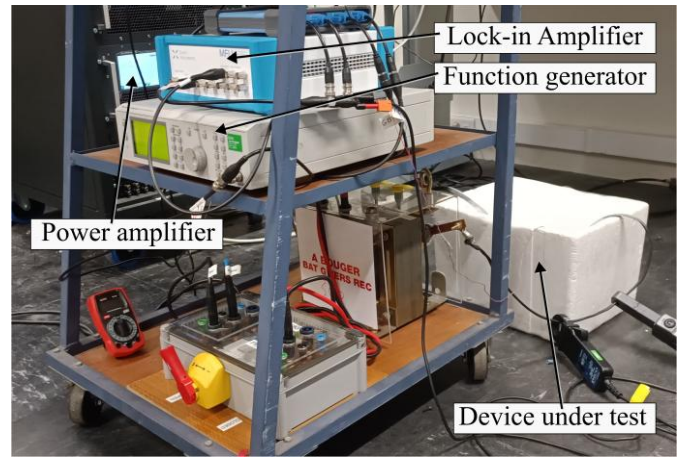

Fig. 5. Plateforme 'PAChinko' pour la mesure des pertes AC.

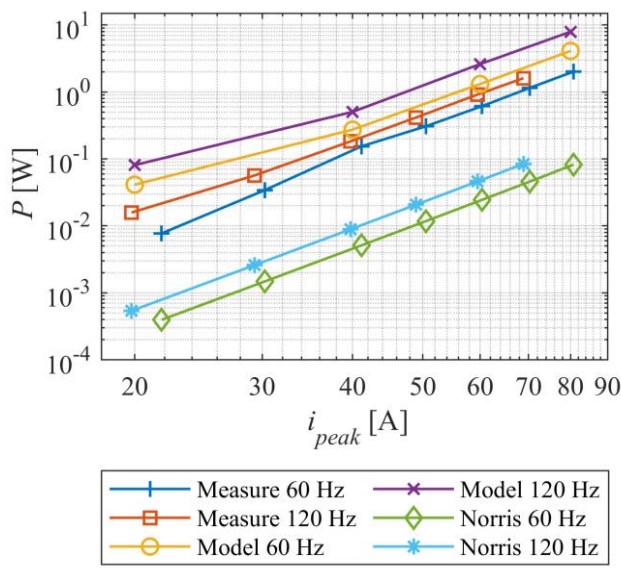

Fig. 6. Pertes AC mesurées en fonction du courant pour 60 et 120 Hz.

### 4. MODELISATION

Pour simuler la bobine, un modèle 3D par éléments finis (FEM) a été utilisé. La formulation en H homogénéisée a été implémentée dans le module PDE du logiciel COMSOL Multiphysics (Fig. 7) [13]. En raison de la symétrie, seulement un quart de la bobine est modélisé.

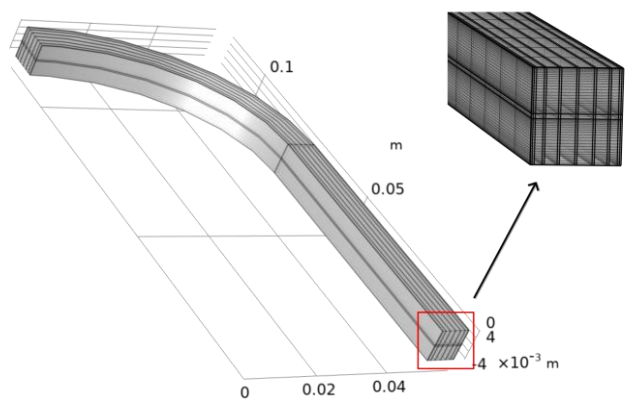

Fig. 7. Modèle éléments finis de la DRC : géométrie et mesh.

En prenant comme données d'entrée du modèle éléments finis les informations du tableau 1, et en modulant la densité de courant critique pour tenir compte du champ magnétique local en utilisant les données [14], la caractéristique *IV* ainsi que les pertes AC de la DRC ont été estimées. Les résultats ont été inclus dans les figures 4 et 6, pour comparaison avec les résultats expérimentaux.

Concernant la caractéristique *IV*, la simulation est en accord avec la mesure. Ceci suggère que le ruban n'a pas été endommagé lors du bobinage et que la diminution du courant critique de la DRC est liée à l'action du champ magnétique propre de la bobine, conformément à ce qui est observé dans d'autres études [12].

Concernant les pertes AC, on observe une différence entre les résultats du modèle et les mesures. Cette différence s'explique par plusieurs facteurs : (i) les propriétés des matériaux sont simplifiées, par exemple l'inhomogénéité du courant critique le long du ruban n'est pas prise en compte ; (ii) les effets thermiques sont omis, notamment l'élévation de température locale en AC, (iii) la densité du maillage et les paramètres du solveur peuvent influencer la précision numérique.

Sur la figure 6, on ajoute les pertes AC calculées à l'aide de la formule analytique de Norris [15]. On observe une différence importante entre les résultats expérimentaux et les prédictions de la formule de Norris. Cette différence vient du fait que le modèle de Norris considère un ruban isolé infiniment long en champ propre, alors que les spires d'une bobine sont courbées et subissent l'influence du champ magnétique local généré par les autres spires.

### 5. CONCLUSION

Nous avons réalisé, caractérisé et modélisé une bobine supraconductrice à haute température critique en ReBCO de type double racetrack. La réalisation est faite à l'aide d'une machine à bobiner spécialement mise au point en interne. La caractérisation comprend la mesure de la caractéristique *IV* et la mesure des pertes en courant alternatif (pertes AC), à l'aide des plateformes de mesure du laboratoire GeePs. La modélisation utilise un modèle éléments finis 3D implémenté dans COMSOL Multiphysics. Les comparaisons réalisées entre les mesures et le modèle montrent une bonne prédiction du courant critique de la bobine, mais révèlent un écart pour les pertes AC. Ils démontrent néanmoins que la bobine est fonctionnelle et que le ruban n'a pas été dégradé lors du bobinage. Ces compétences dans la fabrication et la caractérisation des bobines HTS seront mises à profit pour le développement d'un transformateur supraconducteur.

### 6. REMERCIEMENTS